\documentclass[prd,twocolumn,superscriptaddress,altaffilletter,amssymb,showpacs,nofootinbib]{revtex4}
\usepackage[dvips]{graphicx}
\usepackage{amsmath}
\newcommand{\be}{\begin{equation}}
\newcommand{\ee}{\end{equation}}
\newcommand{\bea}{\begin{eqnarray}}
\newcommand{\eea}{\end{eqnarray}}

\begin{document}

\title{Self-interacting Scalar Field Trapped in a Randall-Sundrum Braneworld: The Dynamical Systems Perspective}

\author{Tam\'e Gonz\'alez}\email{tame@uclv.edu.cu}
\affiliation{Departamento de F\'{\i}sica, Universidad Central de
Las Villas, 54830 Santa Clara, Cuba.} 

\author{Tonatiuh Matos}\email{tmatos@fis.cinvestav.mx}
\affiliation{Departamento de F{\'\i}sica, Centro de Investigaci\'on y de Estudios Avanzados del IPN,\\A.P. 14-740, 07000 M\'exico D.F., M\'exico.} 

\author{Israel Quiros}\email{israel@uclv.edu.cu}
\affiliation{Departamento de F\'{\i}sica, Universidad Central de
Las Villas, 54830 Santa Clara, Cuba.}

\author{Alberto V\'azquez-Gonz\'alez}\email{jvazquez@fis.cinvestav.mx}
\affiliation{Departamento de F{\'\i}sica, Centro de Investigaci\'on
y de Estudios Avanzados del IPN,\\A.P. 14-740, 07000 M\'exico D.F.,
M\'exico.}

\date{\today}
\begin{abstract}
We apply the dynamical systems tools to study the linear dynamics of a self-interacting scalar field trapped on a Randall-Sundrum brane. The simplest kinds of self-interaction potentials are investigated: a) constant potential, and b) exponential potential. It is shown that the dynamics of the Randall-Sundrum model significantly differs from the standard four-dimensional behavior at early times: in all cases of interest the (singular) empty universe is the past attractor for every trajectory in phase space, meanwhile the kinetic energy-dominated solution is always a saddle critical point. The late-time dynamics is not affected by the brane effects.
\end{abstract}

\maketitle

\section{Introduction}

Recent observations from the Wilkinson Microwave Anisotropy Probe (WMAP) \cite{bennet} offer strong supporting evidence in favor of the inflationary paradigm. In the most simple models of this kind, the energy density of the universe is dominated by the potential energy of a single (inflaton) scalar field that slowly rolls down in its self-interaction potential \cite{starobinsky}. Restrictions imposed upon the class of potentials which can lead to realistic inflationary scenarios, are dictated by the slow-roll approximation, and hence, the result is that only sufficiently flat potentials can drive inflation. In order for the potential to be sufficiently flat, these conventional inflationary models should be fine-tuned. This simple picture of the early-time cosmic evolution can be drastically changed if one considers models of inflation inspired in "Unified Theories" like the Super String or M-theory. One of the most appealing models of this kind is the Randall-Sundrum braneworld model of type 2 (RS2) \cite{rs}. In this model a single codimension 1 brane with positive tension is embedded in a five-dimensional anti-de Sitter (AdS) space-time, which is infinite in the direction perpendicular to the brane. In general, the standard model particles are confined to the brane, meanwhile gravitation can propagate in the bulk. In the low-energy limit, due to the curvature of the bulk, the graviton is confined to the brane, and standard (four-dimensional) general relativity laws are recovered.

RS2 braneworld models have an appreciable impact on early universe cosmology, in particular, for the inflationary paradigm. In effect, a distinctive feature of cosmology with a scalar field confined to a RS2 brane is that the expansion rate of the universe differs at high energy from that predicted by standard general relativity. This is due to a term -- quadratic in the energy density -- that produces enhancing of the friction acting on the scalar field. This means that, in RS2 braneworld cosmology, inflation is possible for a wider class of potentials than in standard cosmology \cite{hawkins}. Even potentials that are not sufficiently flat from the point of view of the conventional inflationary paradigm can produce successful inflation. At sufficiently low energies (much less than the brane tension), the standard cosmic behavior is recovered prior to primordial nucleosynthesis scale ($T\sim 1\; MeV$) and a natural exit from inflation ensues as the field accelerates down its potential \cite{5}.\footnote{In this scenario, reheating arises naturally even for potentials without a global minimum and radiation is created through gravitational particle production \cite{6} and/or through curvaton reheating \cite{7}. This last ingredient improves the brane "steep" inflationary picture \cite{8}. Other mechanisms such as preheating, for instance, have also been explored \cite{9}.}

Another interesting feature of this scenario is that the inflaton does not necessarily need to decay; it may survive through the present epoch in the cosmic evolution. Therefore, it may also play the role of the quintessence field, which is a necessary ingredient to explain the current acceleration in the expansion of the universe. Such a unified theoretical framework for the description of both inflaton and quintessence with the help of just one single scalar field has been the target of some works (see for instance Refs. \cite{5,10,11,12}).

Aim of this letter is to study the dynamics of a such scenario -- a self-interacting scalar field trapped on a RS2 brane -- by invoking the dynamical systems tools. These have been proved useful to retrieve significant information about the evolution of a huge class of cosmological models (see for instance the book \cite{coley}).  The simplest self-interaction potentials: a) constant potential, and b) exponential potential, are investigated. The latter one represents a common functional form for self-interaction potentials that can be found in higher-order \cite{witt} or higher-dimensional theories \cite{nd}. These can also arise due to non-perturbative effects \cite{nonp}. In addition to the scalar field we also consider a background fluid trapped on the RS brane.

In a sense, the present research is a generalization of the one reported in \cite{wands}, to include higher-dimensional behaviour (in the present case dictated by the RS dynamics). In consequence, for the exponential potential, our results will include the ones reported in \cite{wands} as a particular case. This is true even if the early-time (high-energy) dynamics does actually differ from the standard (four-dimensional) one, regarding the stability properties of the critical points. This investigation is, also, complementary to the one in reference \cite{plb2008}, where the target was the study of the dynamics of a single scalar field trapped in a Dvali-Gabadadze-Porrati (DGP) brane. In the latter case, however, only the late-time dynamics is modified, while the dynamics of early times is not affected in any essential way. 

Through the paper we use natural units ($8\pi G=8\pi/m_{Pl}^2=\hbar=c=1$).

\section{Phase Space}

The starting point will be the cosmological equations for the RS2 model which, for a Friedmann-Robertson-Walker (FRW) universe with flat spatial sections, can be written in the following way:

\bea &&3H^2=\rho_T(1+\frac{\rho_T}{2\lambda}),\label{friedmann}\\
&&2\dot H=-(1+\frac{\rho_T}{\lambda})(\dot\phi^2+\gamma\rho_m),\label{raycha}\\
&&\dot\rho_m=-3\gamma H\rho_m,\;\;\ddot\phi+\partial_\phi V=-3H\dot\phi.\label{continuity}\eea where $\rho_T=\rho_\phi+\rho_m$ ($\rho_\phi=\dot\phi^2/2+V$, $V$ -- scalar field self-interaction potential) is the total energy content of the brane, $\lambda$ is the brane tension, $\gamma$ is the barotropic index of the background fluid, and the dot accounts for derivative in respect to the cosmic time $t$. Our aim is to write the latter system of second-order (partial) differential equations, as an autonomous system of (first order) ordinary differential equations. For this purpose we introduce the following phase variables:

\be x\equiv\frac{\dot\phi}{\sqrt{6}H},\;\;y\equiv\frac{\sqrt{V}}{\sqrt{3}H},\;\;z\equiv\frac{\rho_T}{3H^2}.\ee After this choice of variables one can realize that

\be \frac{\rho_T}{\lambda}=\frac{2(1-z)}{z},\;\Rightarrow\;0<z\leq 1.\ee This means that the 4D (low-energy) limit of the Randall-Sundrum cosmological equations -- corresponding to the formal limit $\lambda\rightarrow\infty$ -- is associated with the value $z=1$. The high-energy limit $\lambda\rightarrow 0$, on the contrary, corresponds to $z\rightarrow 0$. However, we shall exclude the points with $z=0$ (and their neighborhood) since, in general, at $z=0$ the equations () are undefeined. The critical point associated with $z=0$ has to be analyzed separatelly. In any case the physical meaning of this point in phase space has to be taken with caution due to the high energies associated with it. 

It arises the following constraint:

\be \frac{\rho_m}{3H^2}=z-x^2-y^2.\label{constraint}\ee Since $\rho_m\geq 0$, then $0\leq x^2+y^2\leq z$. We will be focused on expanding FRW universes, so that $y\geq 0$. The resulting phase space for the RS model is the following:

\bea &&\Psi=\{(x,y,z): 0\leq x^2+y^2\leq z,\nonumber\\&&\;\;\;\;\;\;\;\;\;\;\;\;\;\;\;\;\;\;\;\;-1\leq x\leq 1,0\leq y,0<z\leq 1\}.\label{phasespace}\eea

After the above choice of variables of the phase space, the system of cosmological equations (\ref{friedmann}-\ref{continuity}) can be translated into the following autonomous system of ordinary differential equations:

\bea &&x'=-\sqrt\frac{3}{2}(\partial_\phi\ln V)y^2-3x+\nonumber\\&&\;\;\;\;\;\;\;\;\;\;\;\;\;\;\frac{3}{2}\left(\frac{2-z}{z}\right)x[2x^2+\gamma(z-x^2-y^2)],\label{eqx}\\
&&y'=\sqrt\frac{3}{2}(\partial_\phi\ln V)xy+\nonumber\\&&\;\;\;\;\;\;\;\;\;\;\;\;\;\;\frac{3}{2}\left(\frac{2-z}{z}\right)y[2x^2+\gamma(z-x^2-y^2)],\label{eqy}\\
&&z'=3(1-z)[2x^2+\gamma(z-x^2-y^2)],\label{eqz}\eea where the prime denotes derivative with respect to the new time variable $\tau\equiv\ln a$.

It will be helpful to have the parameters of observational importance $\Omega_\phi=\rho_\phi/3H^2$ -- the scalar field dimensionless energy density parameter, and the equation of state (EOS) parameter $\omega_\phi=(\dot\phi^2-2V)/(\dot\phi+2V)$, written in terms of the variables of phase space:

\be \Omega_\phi=x^2+y^2,\;\;\omega_\phi=\frac{x^2-y^2}{x^2+y^2}.\ee Additionally, the decceleration parameter $q=-(1+\dot H/H^2)$:

\be q=-1+\frac{3}{2}\left(\frac{2-z}{z}\right)[2x^2+\gamma(z-x^2-y^2)].\ee

\begin{table*}[tbp]\caption[crit]{Properties of the critical points for the autonomous system (\ref{asode1}).}
\begin{tabular}{@{\hspace{4pt}}c@{\hspace{14pt}}c@{\hspace{14pt}}c@{\hspace{14pt}}c@{\hspace{14pt}}c@{\hspace{14pt}}c@{\hspace{14pt}}c@{\hspace{14pt}}c}
\hline\hline\\[-0.3cm]
$P_i$ &$x$&$y$&$z$&Existence& $\Omega_\phi$& $\omega_\phi$& $q$\\[0.1cm]
\hline\\[-0.2cm]
$P_1$& $0$&$0$&$1$ & Always ($\forall\gamma\in[0,2]$) & $0$&undefined& $(3\gamma-2)/2$\\[0.2cm]
$P_2^\pm$& $\pm 1$&$0$&$1$ & " & $1$&$1$&$2$\\[0.2cm]
$P_3$& $0$&$1$&$1$ & " & $1$&$-1$& $-1$\\[0.2cm]
$P_4$& $0$&$\sqrt z$&$z\in]0,1]$ & " & $z$&$-1$&
$-1$\\[0.4cm]

\hline \hline
\end{tabular}\label{tab1}
\end{table*}
\begin{table*}[tbp]\caption[eigenv]{Eigenvalues for the critical points in table \ref{tab1}.} \begin{tabular}{@{\hspace{4pt}}c@{\hspace{14pt}}c@{\hspace{14pt}}c@{\hspace{14pt}}c@{\hspace{14pt}}c@{\hspace{14pt}}c@{\hspace{14pt}}c}
\hline\hline\\[-0.3cm]
$P_i$ &$x$&$y$& $z$& $\lambda_1$& $\lambda_2$& $\lambda_3$\\[0.1cm]\hline\\[-0.2cm]
$P_1$& $0$&$0$& $1$& $-\frac{3}{2}(2-\gamma)$& $-3\gamma$& $3\gamma/2$\\[0.2cm]
$P_2^\pm$& $\pm 1$&$0$& $1$& $3(2-\gamma)$& $-6$&$3$\\[0.2cm]
$P_3$& $0$&$1$& $1$& $-3$& $-3\gamma$& $0$\\[0.2cm]
$P_4$& $0$&$\sqrt z$& $z\in]0,1]$& $-3$& $-3\gamma$& $0$\\[0.4cm]
\hline \hline
\end{tabular}\label{tab2}
\end{table*}

\section{The Critical Points}

As in \cite{plb2008} we study here two concrete yet generic cases. First we consider a constant potential $V=V_0$, and then we study the exponential potential $V=V_0\exp{(-\alpha\phi)}$. For more general classes of potentials we have to rely on a quite different approach (see, for instance reference \cite{llq}). This will be the subject of future research.

\begin{figure}[t!]
\begin{center}
\includegraphics[width=4cm,height=3.5cm]{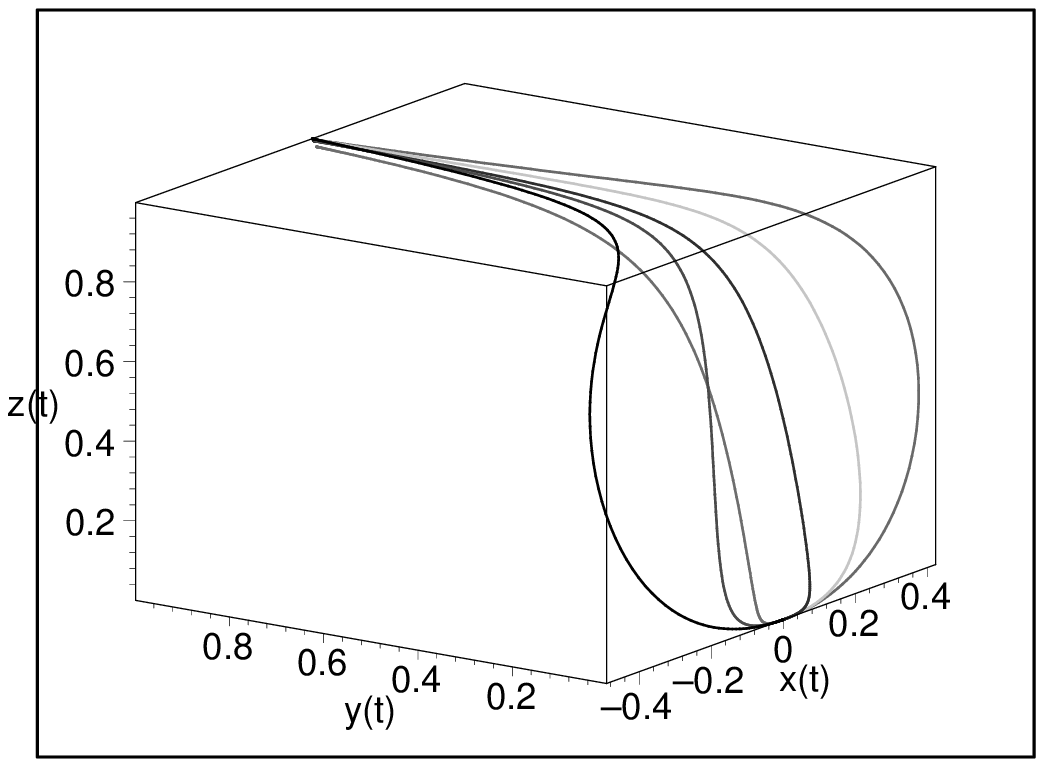}
\includegraphics[width=4cm,height=3.5cm]{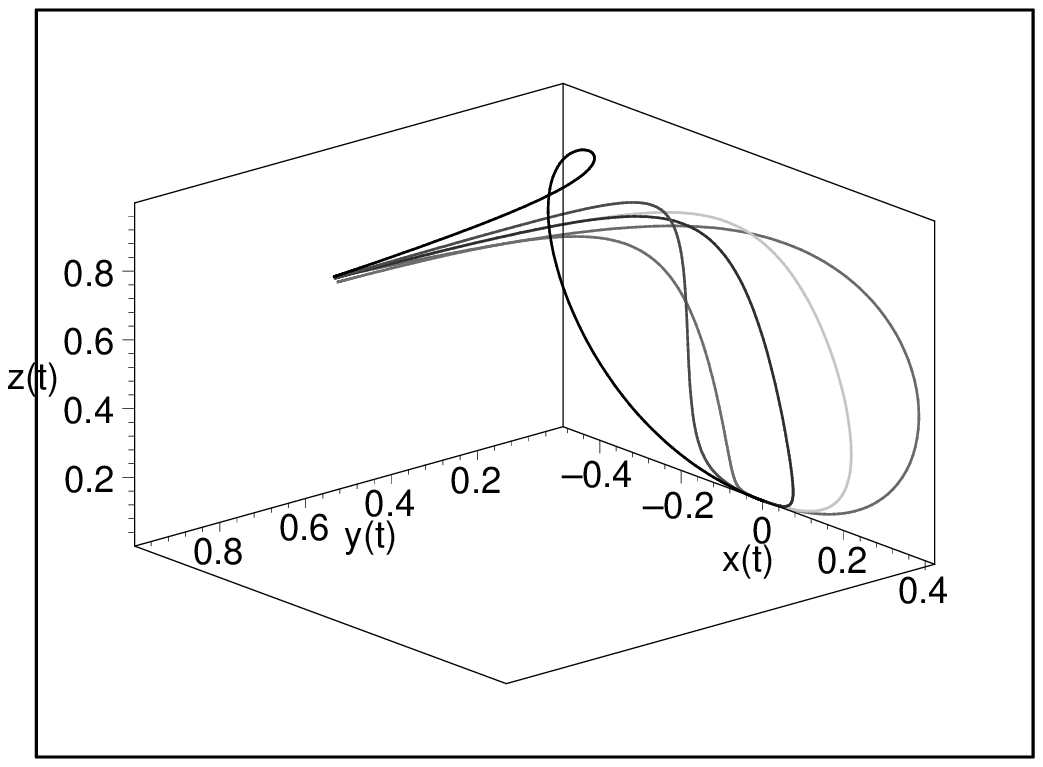}
\includegraphics[width=4cm,height=3.5cm]{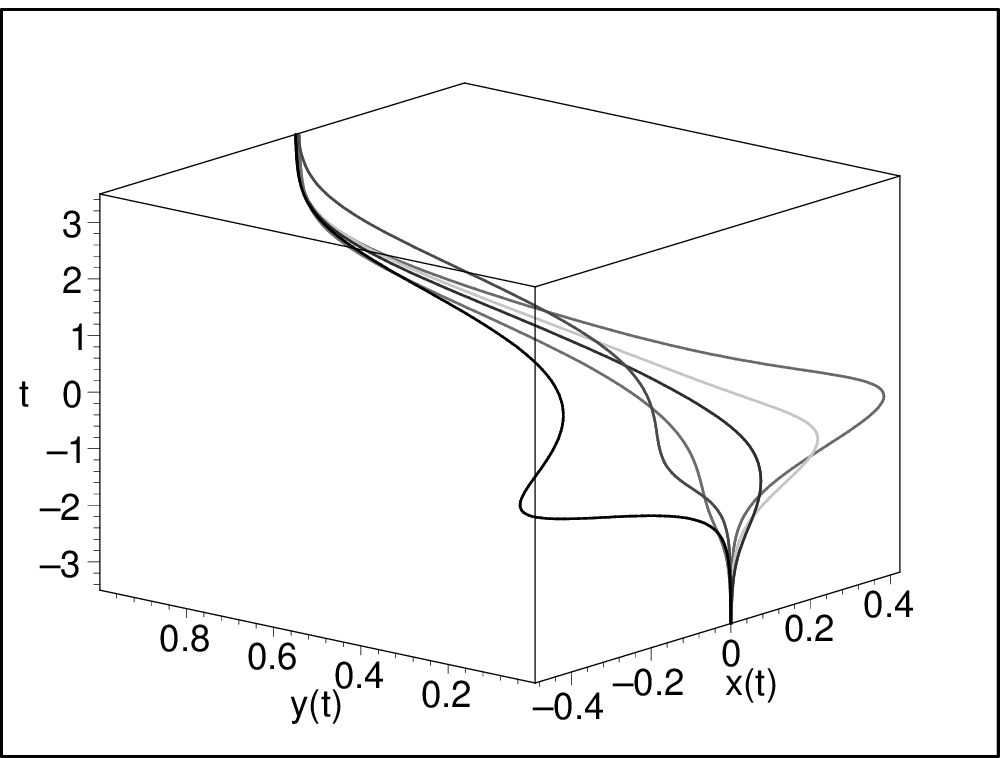}
\includegraphics[width=4cm,height=3.5cm]{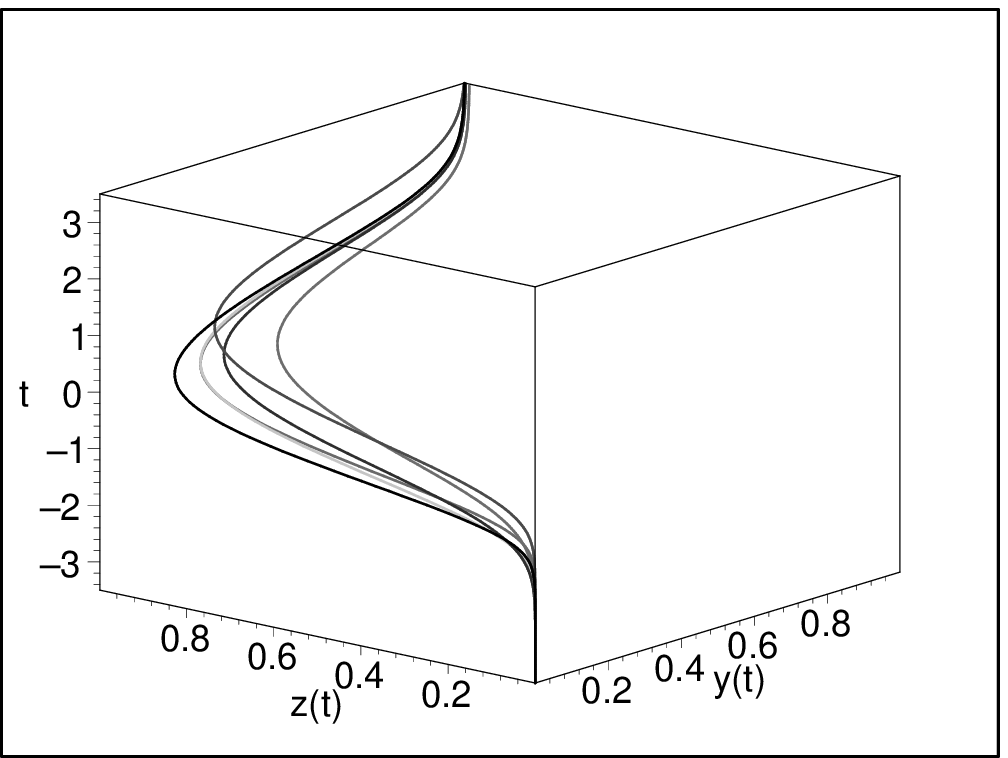}
\vspace{0.3cm}
\caption{Trajectories in phase space for given initial conditions for the constant potential $V=V_0$ (upper part of the figure: view from top to botton in the left panel, and from botton to top in the right panel). The flux in time $\tau$ is shown in the lower part of the figure. At late times trajectories in phase space approach to the plane $(x,y,1)$ that is associated with standard 4D behavior. The late-time (future) attractor is the inflationary de Sitter-FRW universe.}\label{fig1}
\end{center}
\end{figure}

\subsection{The Constant Potential $V=V_0$}\label{v0}

The autonomous system of ordinary differential equations (\ref{eqx}-\ref{eqz}) simplifies to:

\bea &&x'=-3x+\frac{3}{2}\left(\frac{2-z}{z}\right)x[2x^2+\gamma(z-x^2-y^2)],\nonumber\\
&&y'=\frac{3}{2}\left(\frac{2-z}{z}\right)y[2x^2+\gamma(z-x^2-y^2)],\nonumber\\
&&z'=3(1-z)[2x^2+\gamma(z-x^2-y^2)].\label{asode1}\eea

The critical points of the autonomous system of equations (\ref{asode1}) are summarized in table \ref{tab1}, while table \ref{tab2} shows the eigenvalues of the corresponding linearization (Jacobian) matrices. 

Critical points of the autonomous system (\ref{asode1}) can be divided into two sets according to whether $z\neq 1$ or $z=1$. The latter set of points is associated with four-dimensional behavior, meanwhile for the critical points with $z\neq 1$ five-dimensional effects are present. There is just one critical point with $z\neq 1$ (point $P_4$ in table \ref{tab1}): $(x,y,z)=(0,\sqrt z,z)$. Strictly speaking it is not, in fact, a critical point but a subset of critical points. In this case, since $z=y^2$ ($x=0$), then $\rho_m=0$, while $\rho_T=V$, so that the Friedmann equation takes the form:

\be 3H^2=V\left(1+\frac{V}{2\lambda}\right).\label{infla}\ee This represents the slow-roll Friedmann equation relating the Hubble expansion parameter with the potential of the inflaton field, modified by the presence of the RS brane. For potentials much larger than the brane tension (properly the limit when the effects coming from the RS brane are appreciable): $V\gg\lambda\;\Rightarrow\;H_{RS}=V/\sqrt{6\lambda}$, so that the early-time (high-energy) expansion rate in the Randall-Sundrum model $H_{RS}$ gets enhanced with respect to the general relativity rate $H_{GR}$:

\be \frac{H_{RS}}{H_{GR}}=\sqrt\frac{V}{2\lambda}.\ee This is the way brane effects fuel early inflation in the RS model. As seen from table \ref{tab2}, one of the eigenvalues of the linearization matrix for $P_4$ vanishes, so that it is a non-hyperbolic point (set of points), and the linear analysis is not conclusive in this case. However the analysis of the trajectories in phase space reveals the true structure of the phase space. In the figure \ref{fig1} it is shown how the trajectories in phase space emerge from the point $S=(x,y,z)=(0,0,0)$ -- the empty (Misner-RS) universe -- meaning that this is the past attractor of the RS cosmological model. This can be verified by expanding the equations (\ref{eqx}-\ref{eqz}) in the neighborhood of $S=(0,0,0)$, i. e., by replacing $y\rightarrow 0+\epsilon$, $z\rightarrow 0+\epsilon$, while keeping $x=0$ ($\epsilon$ -- a very small perturbation). The result is that the perturbation $\epsilon$ increases exponentially with time $\tau$: $\epsilon(\tau)=\epsilon_0\exp{(3\gamma\tau)}$, so that this point is unstable. 

A cautionary note is neccessary: we want to notice that the points with $z=0$ have been removed from the phase space $\Psi$ since, in general, at $z=0$ the autonomous system of equations (\ref{eqx}-\ref{eqz}) blows up due to our choice of phase space variables. For that reason the point $S$ does not appear in table \ref{tab1}.

There are three additional isolated critical points corresponding to the autonomous system (\ref{asode1}), which are associated with the standard 4D limit of the theory $z=1$. The first of these critical points is the matter-dominated solution: $3H^2=\rho_m$ (point $P_1=(0,0,1)$ in table \ref{tab1}). As seen from table \ref{tab2} it is always a saddle point in phase space. The second one: $P_2^\pm=(\pm 1,0,0)$, is the kinetic energy-dominated solution ($3H^2=\dot\phi^2/2$), which is also a saddle point in phase space. At first sight this is an unexpected result since, in standard general relativity, the kinetic energy-dominated solution is always a source (past attractor) in phase space. However, a closer inspection of the RS model reveals that this point belongs in the high-energy phase, which is the one that is modified by the brane contribution, hence it is not such an unexpected result: five-dimensional contributions modify the structure of the phase space at high-energies. 

The late-time attractor is the inflationary de Sitter-FRW solution $3H^2=V$, which corresponds to the point $P_3=(0,1,1)$ in table \ref{tab1} (see also the figure \ref{fig1}).

\begin{table*}[tbp]\caption[crit1]{Critical points for the autonomous system (\ref{asode2}).}
\begin{tabular}{@{\hspace{4pt}}c@{\hspace{14pt}}c@{\hspace{14pt}}c@{\hspace{14pt}}c@{\hspace{14pt}}c@{\hspace{14pt}}c@{\hspace{14pt}}c@{\hspace{14pt}}c}
\hline\hline\\[-0.3cm]
$P_i$ &$x$&$y$&$z$&Existence& $\Omega_\phi$& $\omega_\phi$& $q$\\[0.1cm]\hline\\[-0.2cm]
$P_1$& $0$&$0$&$1$ & All $\alpha$, $\gamma$ & $0$ &undefined&$-1+\frac{3\gamma}{2}$\\[0.2cm]
$P_2^\pm$& $\pm 1$&$0$&$1$ & All $\alpha$, $\gamma$ &$1$&$1$& $2$\\[0.2cm]
$P_3$& $\frac{\alpha}{\sqrt 6}$&$\sqrt{1-\frac{\alpha^2}{6}}$&$1$ & $\alpha^2<6$& $1$&$\frac{\alpha^2}{3}-1$& $\frac{\alpha^2}{2}-1$\\[0.2cm]
$P_4$&$\sqrt\frac{3}{2}\frac{\gamma}{\alpha}$&$\sqrt\frac{3(2-\gamma)\gamma}{2\alpha^2}$&$1$ & $\alpha^2>3\gamma$ &$\frac{3\gamma}{\alpha^2}$&$\gamma-1$& $-1+\frac{3\gamma}{2}$\\[0.4cm]\hline \hline
\end{tabular}\label{tab3}
\end{table*}

\begin{table*}[tbp]\caption[eigenv]{Eigenvalues for the critical points in table \ref{tab3}. In the row corresponding to the last critical point $P_4$, for compactness, we have introduced the notation $\Pi\equiv\sqrt{(2-\gamma)(24(\gamma/\alpha)^2-9\gamma+2)}$.} \begin{tabular}{@{\hspace{4pt}}c@{\hspace{14pt}}c@{\hspace{14pt}}c@{\hspace{14pt}}c@{\hspace{14pt}}c@{\hspace{14pt}}c@{\hspace{14pt}}c}
\hline\hline\\[-0.3cm]
$P_i$ &$x$&$y$& $z$& $\lambda_1$& $\lambda_2$& $\lambda_3$\\[0.1cm]\hline\\[-0.2cm]
$P_1$& $0$&$0$& $1$& $-\frac{3}{2}(2-\gamma)$& $-3\gamma$& $3\gamma/2$\\[0.2cm]
$P_2^\pm$& $\pm 1$&$0$& $1$& $\sqrt\frac{3}{2}(\alpha-\sqrt{6})$& $-6$&$3(2-\gamma)$\\[0.2cm]
$P_3$& $\frac{\alpha}{\sqrt 6}$&$\sqrt{1-\frac{\alpha^2}{6}}$& $1$& $\alpha^2-3\gamma$& $\frac{\alpha^2-6}{2}$& $-\alpha^2$\\[0.2cm]
$P_4$& $\sqrt\frac{3}{2}\frac{\gamma}{\alpha}$&$\sqrt\frac{3(2-\gamma)\gamma}{2\alpha^2}$& $1$& $-\frac{3}{4}[(2-\gamma)+\Pi]$& $-\frac{3}{4}[(2-\gamma)-\Pi]$& $-3\gamma$\\[0.4cm]
\hline \hline
\end{tabular}\label{tab4}
\end{table*}

\begin{figure}[t!]
\begin{center}
\includegraphics[width=4.0cm,height=3.5cm]{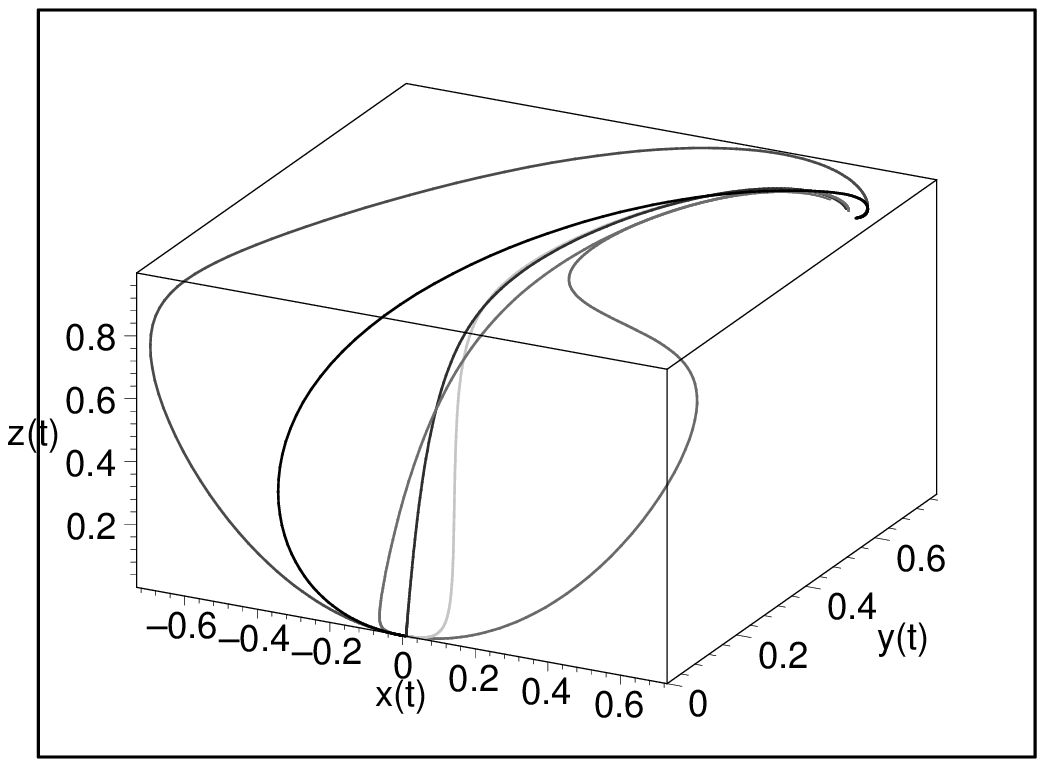}
\includegraphics[width=4.0cm,height=3.5cm]{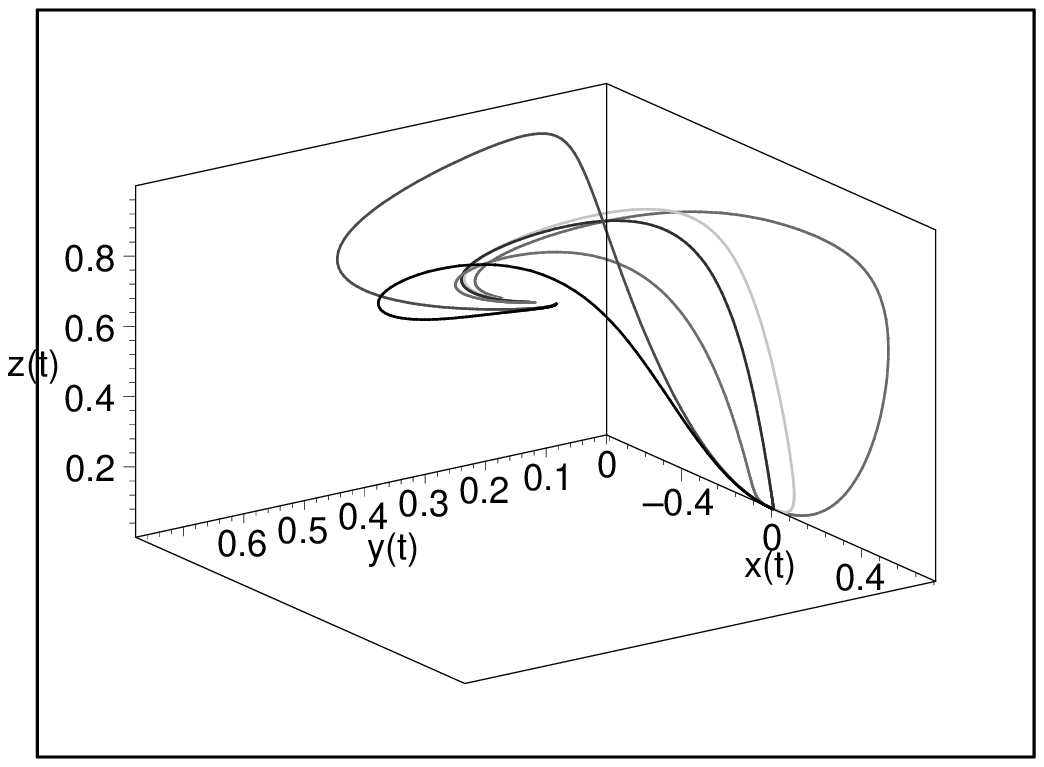}
\includegraphics[width=4.0cm,height=3.5cm]{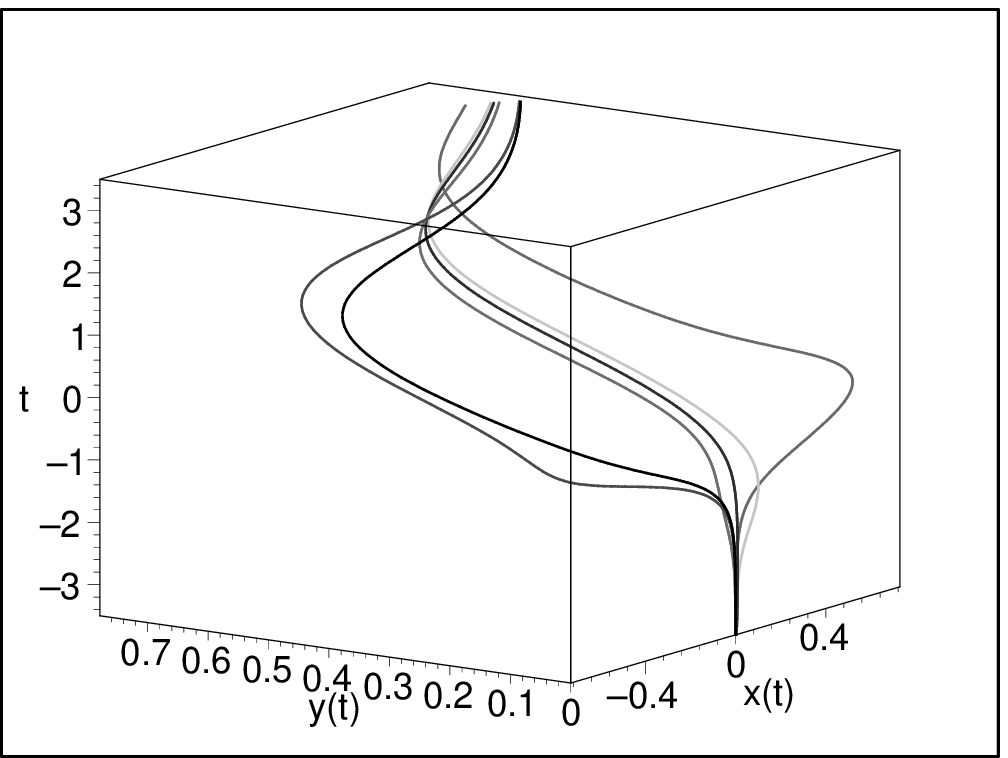}
\includegraphics[width=4.0cm,height=3.5cm]{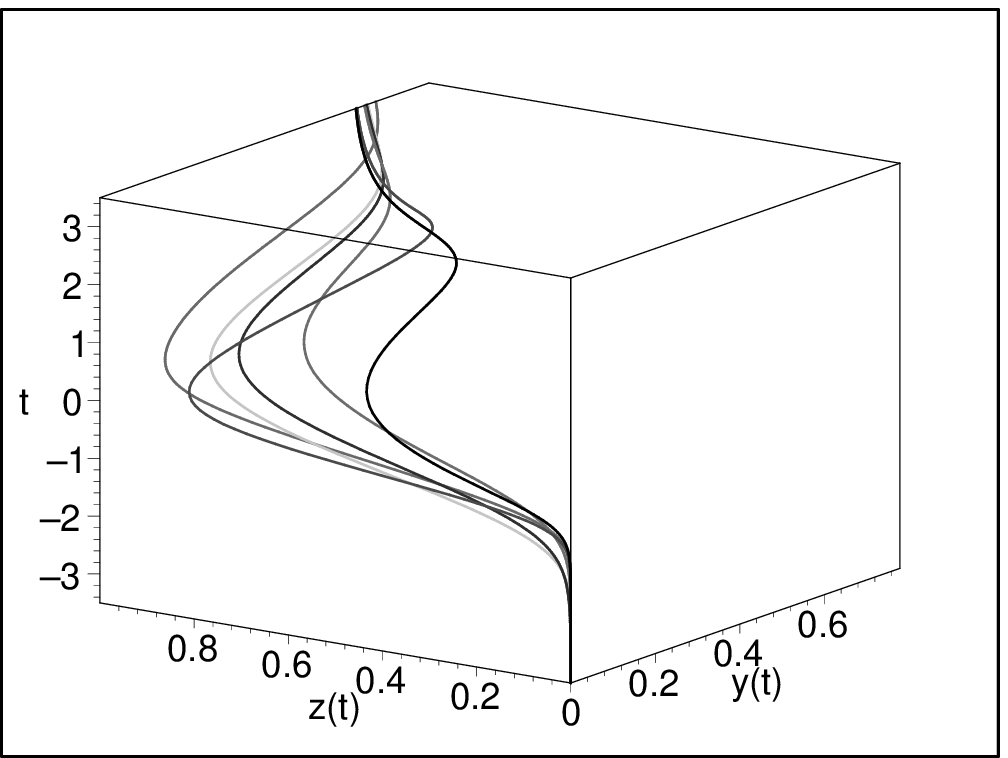}
\vspace{0.3cm}
\bigskip
\caption{Phase trajectories for given initial data, for the exponential potential $V=V_0\exp{(-\alpha\phi)}$ (upper part of the figure: view from top to botton in the left panel, and from botton to top in the right panel). The corresponding flux in time $\tau$, is shown in the panels in the lower part of the figure. The exponent has been chosen to be $\alpha=2$ so that the scalar field-dominated solution is the late-time attractor.}\label{fig2}
\end{center}
\end{figure}

\begin{figure}[ht!]
\begin{center}
\includegraphics[width=4.0cm,height=3.5cm]{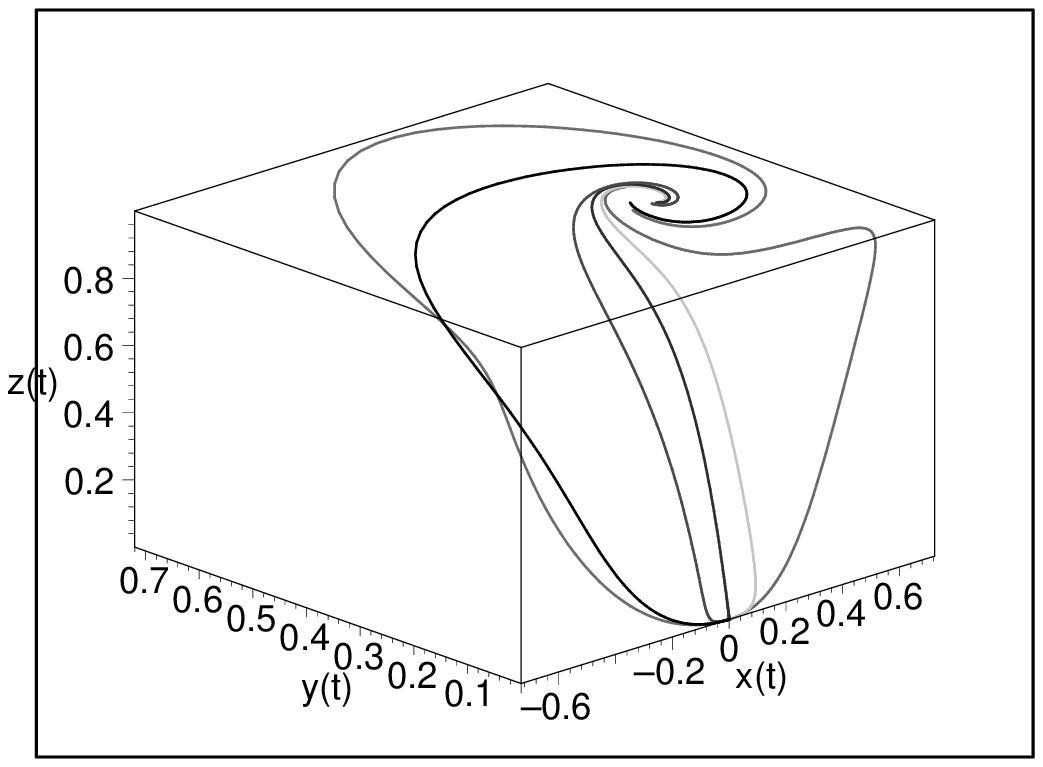}
\includegraphics[width=4.0cm,height=3.5cm]{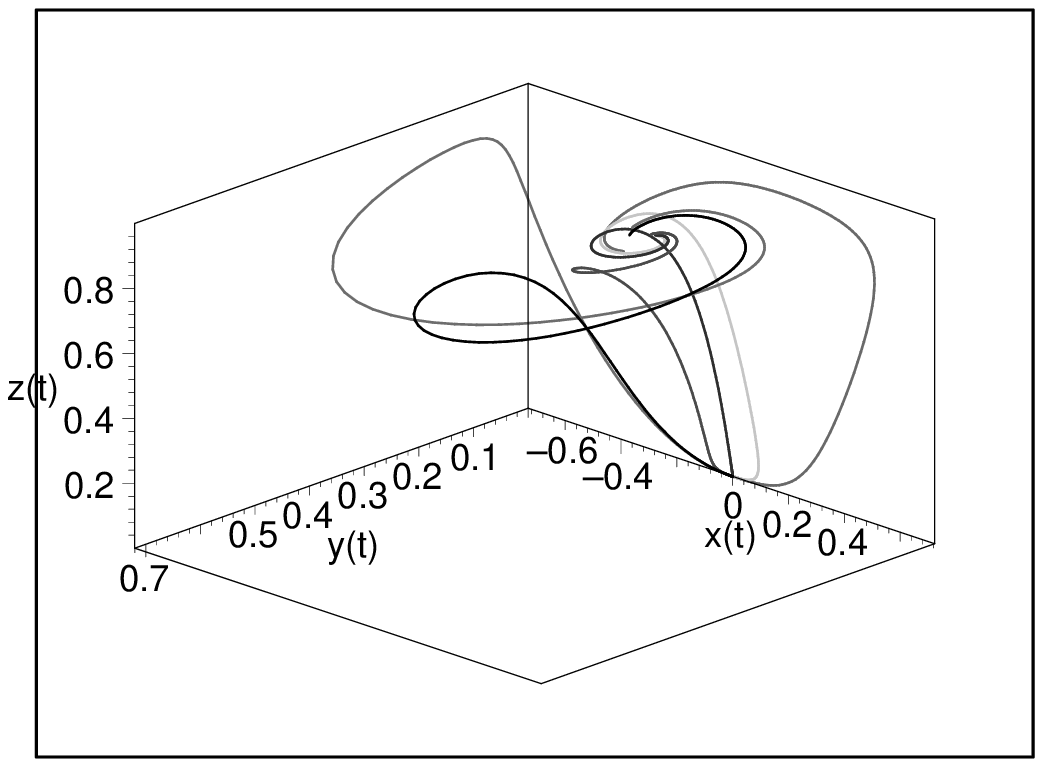}
\includegraphics[width=4.0cm,height=3.5cm]{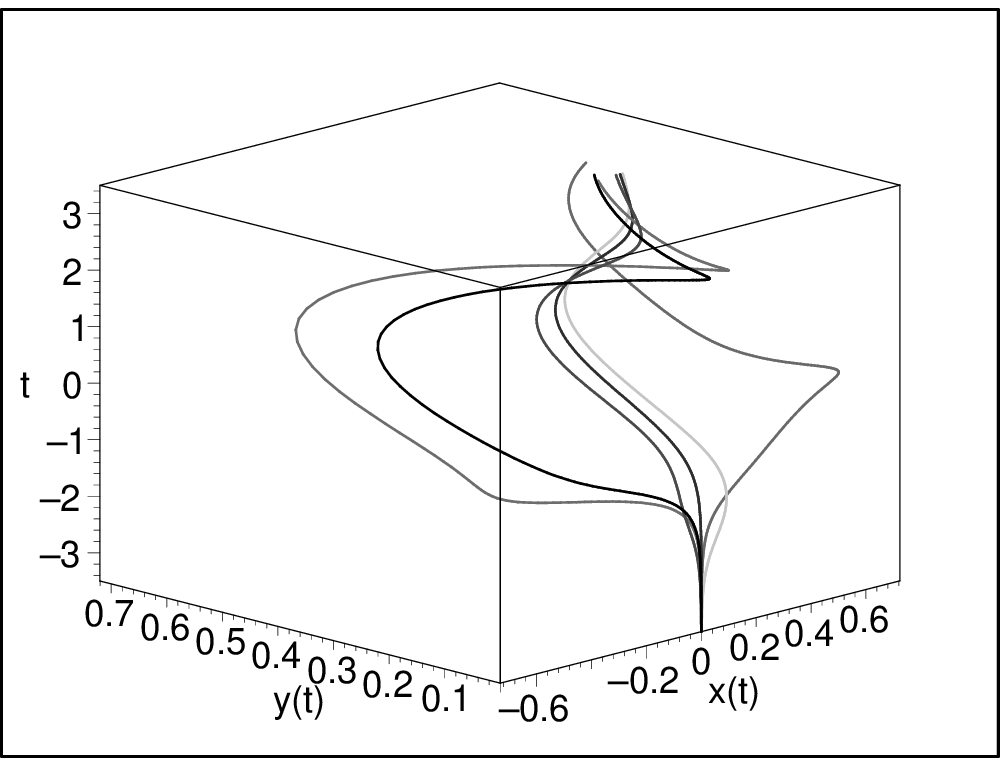}
\includegraphics[width=4.0cm,height=3.5cm]{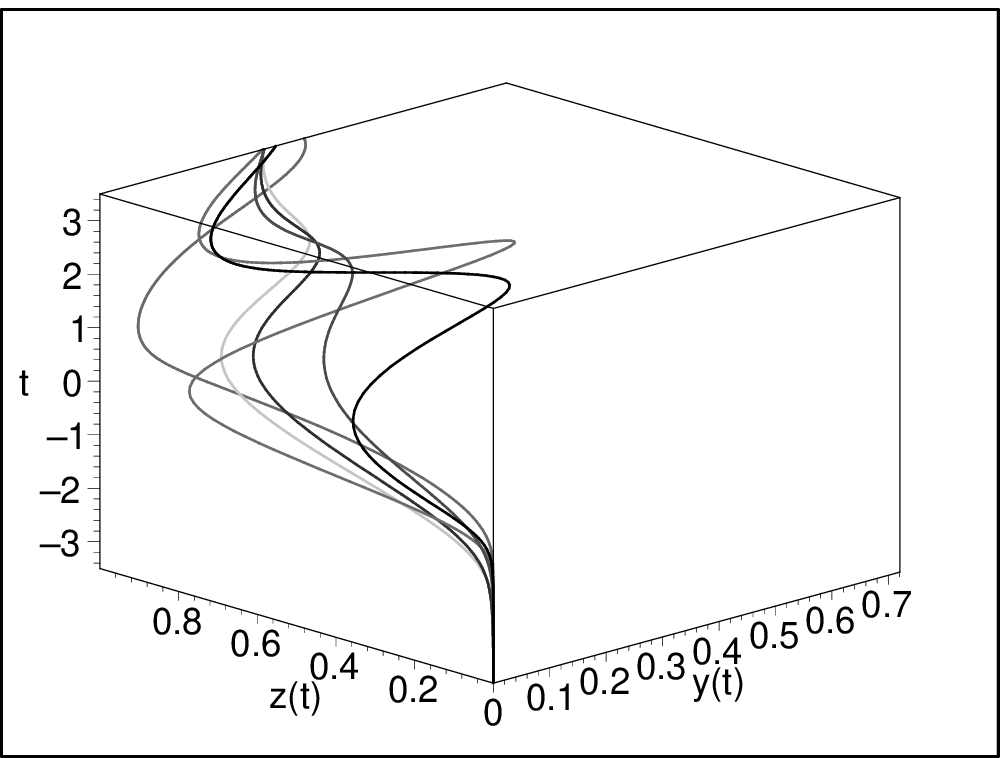}
\vspace{0.3cm}
\bigskip
\caption{Trajectories in phase space for given initial data, for the exponential potential $V=V_0\exp{(-\alpha\phi)}$ (upper part of the figure), and the corresponding flux in time $\tau$ (lower part of the figure). The exponent now has been set $\alpha=4$ so that the matter-scaling solution is the late-time attractor in this case.}\label{fig3}
\end{center}
\end{figure}

\subsection{The exponential Potential $V=V_0\exp{(-\alpha\phi)}$}\label{expot}

The autonomous system of ordinary differential equations (\ref{eqx}-\ref{eqz}) takes the following form:

\bea &&x'=\sqrt\frac{3}{2}\alpha y^2-3x+\nonumber\\&&\;\;\;\;\;\;\;\;\;\;\;\;\;\;\;\frac{3}{2}\left(\frac{2-z}{z}\right)x[2x^2+\gamma(z-x^2-y^2)],\nonumber\\
&&y'=\sqrt\frac{3}{2}\alpha xy+\frac{3}{2}\left(\frac{2-z}{z}\right)y[2x^2+\gamma(z-x^2-y^2)],\nonumber\\
&&z'=3(1-z)[2x^2+\gamma(z-x^2-y^2)].\label{asode2}\eea 

The critical points of (\ref{asode2}) are summarized in table \ref{tab3}, while the corresponding eigenvalues are shown in table \ref{tab4}. In this case there is only one critical point that can be associated with higher-dimensional behavior ($z\neq 1$): $S=(x,y,z)=(0,0,0)$, instead of a whole set of points as in the former case. Actually, at the critical point $x'=y'=z'=0$, therefore, from the third equation in (\ref{asode2}) it follows that 

\be z'=0\;\;\Rightarrow\;\;z=\left(\frac{\gamma-2}{\gamma}\right) x^2+y^2,\ee meanwhile, simultaneously, 

\bea x'=0\;\;\Rightarrow\;\;x=\frac{\alpha}{\sqrt 6}y^2-3x,\nonumber\\y'=-\sqrt\frac{3}{2}\alpha x y=0\;\;\Rightarrow\;\;x=y=z=0.\eea We have to recall, however, that the points with coordinate $z=0$ have been removed from the phase space. This is due to technical reasons related with our choice of coordinates. For that reason, as in the former case, the point $S$ does not appear in table \ref{tab2}.

To understand what happens at the point $(x,y,z)=(0,0,0)$, it is recommended to expand the equations (\ref{asode2}) in its neighborhood, i. e., to replace $x\rightarrow 0+\epsilon$, $y\rightarrow 0+\epsilon$. As before one gets that the (homogeneous) perturbation $\epsilon$ grows exponentially in time $\tau$: $\epsilon(\tau)=\epsilon_0\exp{(3\gamma\tau)}$, so that this point is a source (past attractor) in phase space. 

\begin{figure}[t]
\begin{center}
\includegraphics[width=5cm,height=4.5cm]{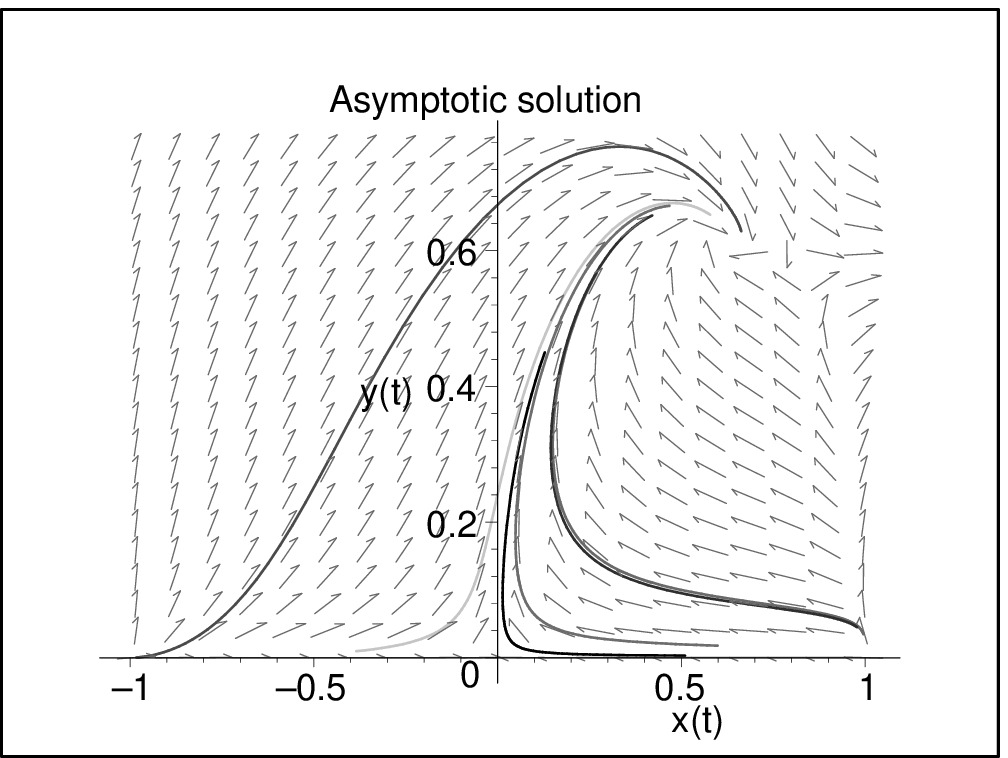}
\includegraphics[width=5cm,height=4.5cm]{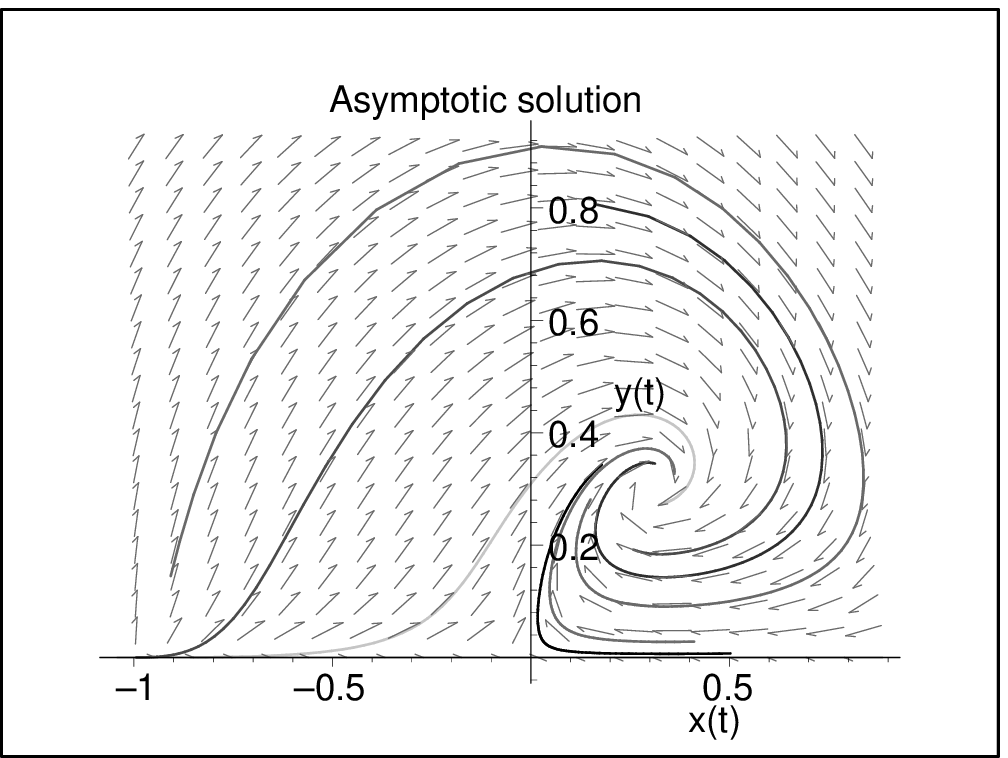}
\vspace{0.3cm}
\bigskip
\caption{Projections of the phase trajectories onto the plane $(x,y,z=1)$ -- associated with low-energy (late-time) behavior -- for the exponential potential $V=V_0\exp{(-\alpha\phi)}$ ($\alpha=2$ in the upper figure, while $\alpha=4$ in the lower figure). Note that, although the kinetic energy-dominated/matter-scaling solution (upper/lower part of the figure respectively) appears to be the past attractor, it is just a "mirage" -- an effect of the projection -- due to the ignorance of the brane effects.}\label{fig4}
\end{center}
\end{figure}

There are four additional (isolated) critical points which are associated with the 4D limit. A complete study of these critical points (see table \ref{tab3}), for the standard 4D case ($z=1$), can be found in the well known reference \cite{wands}. However, at early times, the results of \cite{wands} differ a little bit from the ones reported here, due to the modifications produced by the RS brane effects on the cosmic dynamics. Our main results can be summarized as follows (see tables \ref{tab3} and \ref{tab4}):

\begin{itemize}

\item{$\forall\alpha,\gamma$ ($0\leq\gamma\leq 2$);} The past attractor in the phase space is the empty universe (point $S$ that does not appear in table \ref{tab3}) (figure \ref{fig2}). The matter-dominated solution (point $P_1$ in table \ref{tab3}), and the kinetic energy-dominated solutions (points $P_2^\pm$), are always saddle points in phase space.\footnote{As it has been properly noted before, due to the influence of the brane effects, the early-times dynamics is modified with respect to the standard 4D general relativity situation.}

\item{$\alpha^2<3\gamma$;} The scalar field-dominated solution (point $P_4$) is the late-time attractor (see figure \ref{fig2}).

\item{$3\gamma<\alpha^2<6$;} The scalar field-dominated solution $P_3$ is a saddle point. Whenever $24(\gamma/\alpha)^2$ is bigger/lower than $9\gamma-2$, the matter-scaling solution (point $P_4$ in table \ref{tab3}) is a stable node/spiral (see figures \ref{fig3} and \ref{fig4}).

\item{$\alpha^2>6$;} The matter-scaling solution $P_4$ is a stable spiral (see figures \ref{fig3} and \ref{fig4}).

\end{itemize}

\section{Results and Discussion}

From the analysis in the former sections, the following important results can be summarized:

\begin{itemize}

\item For the constant potential $V=V_0$, early inflation fuelled by the potential energy of the scalar field trapped in the Randall-Sundrum brane (equation (\ref{infla})), is a (non-isolated) critical point in phase space.\footnote{As already noted, it is in fact a critical set or, strictly speaking, a set of critical points: $S=(0,\sqrt z,z)$, $z\in]0,1]$.} The late-time attractor is the (inflationary) de Sitter space.

\item For the exponential potential $V=V_0\exp{(-\alpha\phi)}$, early inflation is not even a critical point. This means that, for this potential, inflation is not generic even if the higher-dimensional effects in the Randall-Sundrum scenario favor early inflation. Depending on the values of the free parameters $\alpha$ and $\gamma$, the late-time attractor can be either the scalar field-dominated solution ($\alpha^2<3\gamma$, inflationary if $\alpha^2<2$), or the matter-scaling solution ($\alpha^2>3\gamma$).

\item For both potentials the past (high-energy) attractor is the (singular) empty universe ($z=0\;\Rightarrow\;\rho_T=0$). However, given the high energies implied, the physical significance of this point is suspicious. The kinetic energy-dominated solution is always a saddle point in phase space.

\end{itemize}

The latter result marks a difference with respect to the well known results within general relativity plus a minimally coupled scalar field \cite{wands}, according to which the kinetic energy-dominated solution is the past attractor for $V=V_0$, while, for the exponential potential it can be either a past attractor, or a saddle point. The origin of this difference is in the high energy modificatios of the dynamics caused by the brane effects in the Randall-Sundrum scenario, leading to the empty universe being always the past attractor.\footnote{As already noted, this critical point has to be taken with some care due to the very high energies associated with it. One has to be sure of the validity of the Randall-Sundrum model at such large energies.}

Another interesting result has to do with the fact that, for the constant potential $V=V_0$, the early-time inflation driven by the potential energy of the scalar field -- properly the cosmological constant -- represents a critical subset in the phase space of the RS model: $S=(0,\sqrt z,z)$,

\be z=y^2\;\Rightarrow\;3H^2=V_0(1+\frac{V_0}{2\lambda}).\label{x}\ee For sufficiently large values of the brane cosmological constant $\Lambda=V_0\gg\lambda$, the Hubble expansion rate is enhanced with respect to the standard general relativity (GR) rate: $H=V_0/\sqrt{6\lambda}$. 

This subset of critical points could be associated with early inflation. The fact that this solution is a subset of critical points in phase space, means that, for $V=V_0$, early inflation described by equation (\ref{x}), is quite a generic property of the cosmological RS model. For the exponential potential, on the contrary, early inflation is not even a critical point in the phase space, meaning that the genericity of inflation is dependent on the kind of self-interaction potential under consideration.

The phase space pictures in the figures \ref{fig1}, \ref{fig2}, and \ref{fig3}, reveal the actual behavior of the RS dynamics: trajectories in phase space depart from the (singular) empty universe, possibly related with the initial (Big-Bang) singularity, and, at late times, approach to the plane $(x,y,1)$, which is associated with standard four-dimensional behavior.

\section{Conclusion}

The dynamics of RS models departs from the one in general relativity at high energies (early times). In particular, a dynamical systems approach to the subject reveals that the kinetic energy-dominated solution, which is usually the past attractor for every trajectory in the phase space, is replaced by the empty universe. The kinetic energy-dominated phase represents always a saddle point. At late times gravity is effectivelly confined to the brane so that the late-time dynamics is not modified with respect to the standard general relativity case.

For the constant potential case there are two critical points -- properly a subset of non-isolated critical points, plus an isolated critical point -- that are linked with inflationary behavior. This entails a good possibility for this model to account for an unified description of early inflation and of the late-time accelerated expansion of the universe, with the help of a single scalar field. This possibility, whithin the frame of the Randall-Sundrum model, has been studied, in other contexts, for instance, in reference \cite{prd2005}. However, for the exponential potential the situation is quite different: there are no critical points that could be associated with early inflation. Hence, the possibility for an unified description of inflation and quintessence within RS braneworld model is not as generic as thought. 

It can be of insterest to investigate the present scenario for arbitrary self-interaction potentials, so that we will be able to reach to more generic results. This task would entail a different approach than the one undertaken in this paper,\footnote{See, for instance, the recent reference \cite{chinos}.} so that we leave it for future work.

The authors want to aknowledge very useful discussions with Luis Urena, Alfredo Herrera and Ulises Nucamendi. This work was partly supported by CONACyT M\'exico, under grants 49865-F, 54576-F, 56159-F, 49924-J, and by grant number I0101/131/07 C-234/07, Instituto Avanzado de Cosmologia (IAC) collaboration. T M and I Q aknowledge also the MES of Cuba for partial support of the research.

\end{document}